# Exploring the SARS-CoV-2 virus-host-drug interactome for drug repurposing

Sepideh Sadegh[1,a], Julian Matschinske[1,a], David B. Blumenthal[1,b], Gihanna Galindez[1,b], Tim Kacprowski[1,b], Markus List[1,b], Reza Nasirigerdeh[1,b], Mhaned Oubounyt[1,b], Andreas Pichlmair[2,b], Tim Daniel Rose[3,b], Marisol Salgado-Albarrán[1,4,b], Julian Späth[1,b], Alexey Stukalov[2,b], Nina K. Wenke[1,b], Kevin Yuan[1,b], Josch K. Pauling[3], Jan Baumbach[1,5]

[1] Chair of Experimental Bioinformatics, TUM School of Life Sciences, Technical University of Munich
[2] Institute of Virology, TUM School of Medicine, Technical University of Munich
[3] LipiTUM, Chair of Experimental Bioinformatics, TUM School of Life Sciences, Technical University of Munich
[4] Natural Sciences Department, Universidad Autónoma Metropolitana-Cuajimalpa (UAM-C), Mexico City, 05300, Mexico
[5] Computational Biomedicine lab, Department of Mathematics and Computer Science, University of Southern Denmark

[a] joint first authors
[b] alphabetical order

## Abstract

Coronavirus Disease-2019 (COVID-19) is an infectious disease caused by the SARS-CoV-2 virus. It was first identified in Wuhan, China, and has since spread causing a global pandemic. Various studies have been performed to understand the molecular mechanisms of viral infection for predicting drug repurposing candidates. However, such information is spread across many publications and it is very time-consuming to access, integrate, explore, and exploit. We developed CoVex, the first interactive online platform for SARS-CoV-2 and SARS-CoV-1 host interactome exploration and drug (target) identification. CoVex integrates 1) experimentally validated virus-human protein interactions, 2) human protein-protein interactions and 3) drug-target interactions. The web interface allows user-friendly visual exploration of the virus-host interactome and implements systems medicine algorithms for network-based prediction of drugs. Thus, CoVex is an important resource, not only to understand the molecular mechanisms involved in SARS-CoV-2 and SARS-CoV-1 pathogenicity, but also in clinical research for the identification and prioritization of candidate therapeutics. We apply CoVex to investigate recent hypotheses on a systems biology level and to systematically explore the molecular mechanisms driving the virus' life cycle. Furthermore, we extract and discuss drug repurposing candidates involved in these mechanisms. CoVex renders COVID-19 drug research systems-medicine-ready by giving the scientific community direct access to network medicine algorithms integrating virus-host-drug interactions. It is available at https://exbio.wzw.tum.de/covex/.

## Introduction

Coronavirus Disease-2019 (COVID-19) is an infectious disease caused by SARS-CoV-2 (severe acute respiratory syndrome coronavirus 2). It was first identified in Wuhan, China and has spread causing an ongoing pandemic [1] with globally 2.4 million confirmed cases and 167 thousand deaths as of April 20, 2020.

Our insights into SARS-CoV-2 infection mechanisms are limited and clinical therapy has largely focused on treating critical symptoms. Therefore, the current pandemic requires fast and freely accessible knowledge to accelerate the development of vaccines, treatments and diagnostic tests. Research data has been collected in several online platforms such as, the COVID-19 Open Research Dataset and the Dimensions COVID-19 collection [2,3]. In addition, existing databases that collect virus information have responded by integrating new SARS-CoV-2 research [4,5].

As vaccine and drug development may take years, drug repurposing is a potent approach that offers new therapeutic options through the identification of alternative uses of already approved drugs [6]. These drugs have previously undergone clinical and safety trials and, hence, accelerate drug development timelines from a decade to a few years or months. Due to the COVID-19 pandemic, numerous research groups around the world have been joining their efforts to identify drugs that can be repurposed to effectively treat COVID-19. Numerous drugs are already part of clinical trials, including remdesivir (a less effective ebola drug), Chloroquine, Hydroxychloroquine (antimalarial drugs), Tocilizumab (rheumatoid arthritis drug), Favipiravir (influenza drug), and Kaletra (a combination of Lopinavir and Ritonavir for treating human immunodeficiency virus HIV-1) [7].

Computational systems and network medicine approaches offer a methodological toolbox required to understand molecular virus-host-drug mechanisms and to predict novel drug targets to attack them [8,9]. Few studies on these mechanisms in SARS-CoV-2 exist. Gordon *et al.* applied affinity purification-mass spectrometry (AP-MS) to reconstruct the SARS-CoV-2-human protein-protein interaction (PPI) network and subsequently employed a chemoinformatics approach to identify potential drugs for repurposing [10]. The data generated from that study is a major advancement in understanding SARS-CoV-2 infection. However, to identify drug candidates, the study mainly considered the direct interactors of the human proteins as putative targets and thus did not take into account the network context of the human interactome. However, viral interactions with human proteins have cascading effects in the human interactome, where key proteins necessary for the viral replication cycle are only indirectly affected. Therefore, downstream host proteins may be additional promising targets for therapeutic intervention but require thorough data integration and mining to be identified (see Supplementary Material for details). Figure 1 illustrates the concept of systems medicine-based drug repurposing specifically for SARS-CoV-2.

Gysi *et al.* integrated the experimentally validated SARS-CoV-2 virus-host interactions with the human interactome and investigated co-morbidity and differences of virus-host interactions across 56 tissues [11]. Furthermore, network medicine analysis was applied to compile a list of drug repurposing candidates that target also indirectly affected proteins in the human interactome. However, the combined number of virus-host, host-host and drug-target interactions goes into the millions such that purely algorithmic approaches to discovering new drug targets and drug repurposing candidates produces a large number of results, many of which lack mechanistic specificity and, hence, are not useful. Thus, to make their results accessible, Gysi *et al.* worked closely together with clinical experts to narrow down the number of predicted repurposable drugs.

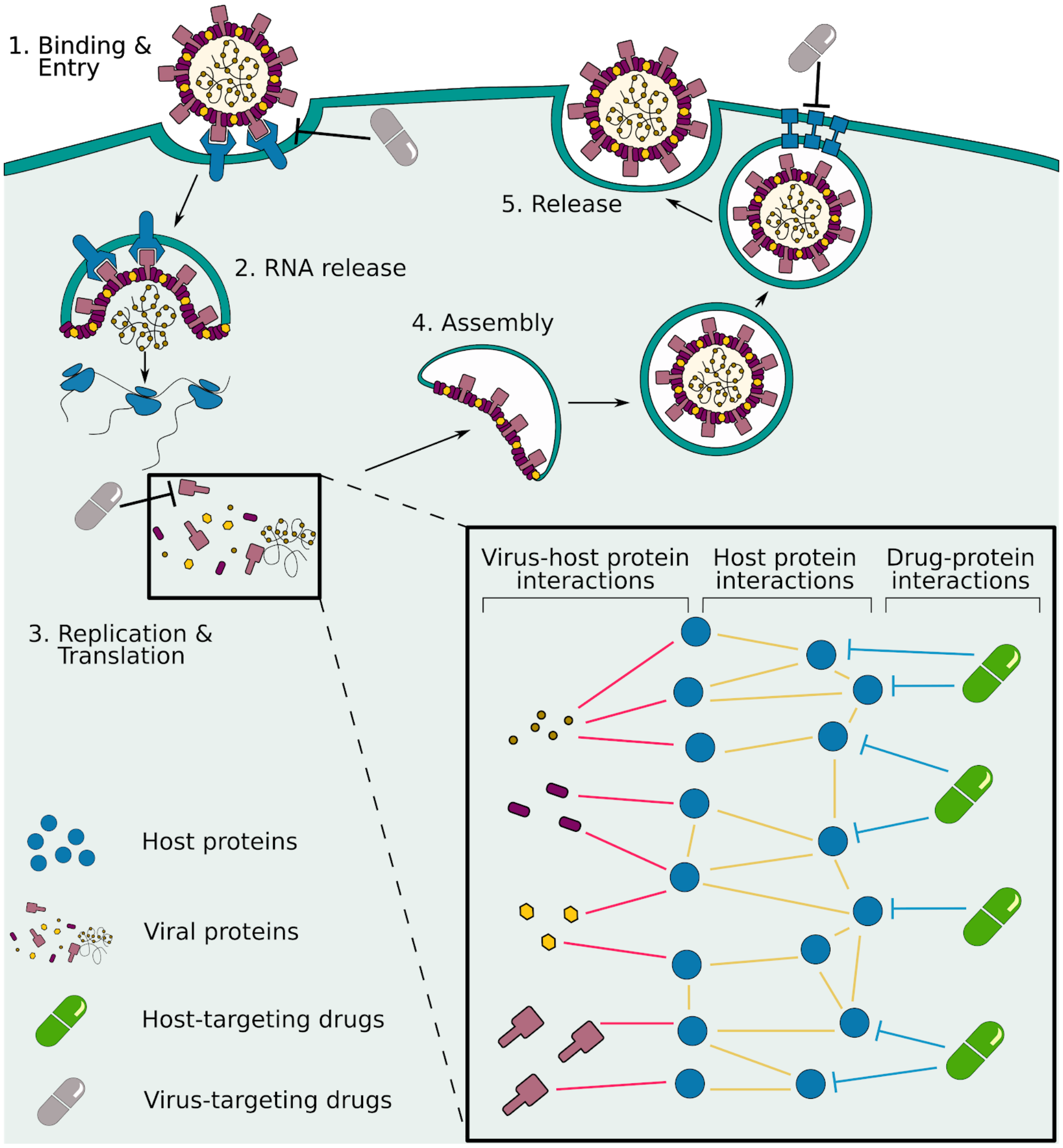

**Figure 1 - The SARS-CoV-2 life cycle and the CoVex systems medicine approach of drug repurposing.** Most antiviral drugs (grey) target the virus proteins or their direct host interactor proteins to inhibit different stages of the viral life cycle. Our rationale, however, is that viral interactions with human host proteins have a cascading effect to hijack and control key proteins necessary for the virus' life cycle. We aim to identify repurposable drug candidates (green) targeting these key host modulators to interfere with virus replication and disease progression following infection. Besides an increased antiviral drug repertoire, targeting host proteins would make it more difficult for the virus (population) to develop resistance mutations.

In order to allow for the interactive integration of expert knowledge about virus replication, immune-related biological processes or drug mechanisms, we have developed the interactive systems and network medicine platform CoVex (CoronaVirus Explorer). It integrates experimental virus-human interaction data for SARS-CoV-2 and SARS-CoV-1 with the human interactome as well as drug information to predict novel drug (target) candidates, and it offers biomedical and clinical researchers interactive and user-friendly access to network medicine algorithms for advanced data mining and hypothesis testing. CoVex follows a human-in-the-loop paradigm and provides an intuitive visualization of virus-host interactions, drug targets, and drugs to enable researchers to examine molecular mechanisms that can be targeted using repurposed drugs. CoVex offers two main actions for which several network medicine algorithms are available: Given a list of user-selected human host proteins, viral proteins, or drugs (referred to as seeds), users can (1) search the human interactome for viable drug targets and (2) identify repurposable drug candidates. In a typical work-flow, these two actions are combined, i.e. starting from a selection of virus or virus-interacting proteins, users mine the interactome for suitable drug targets for which, in turn, suitable drugs are identified. Additionally, users can leverage expert knowledge by uploading a list of proteins or drugs of interest as seeds to guide the analysis. Such seeds could, for instance, be a list of differentially expressed genes, a list of proteins related to a molecular mechanism of interest, or a set of drugs known to be effective.

The remainder of this paper is structured as follows: In the methods section, we first describe the datasets and integration strategy used in CoVex. Next, we introduce the rationales of the systems and network medicine algorithms implemented in CoVex, and briefly describe the overall architecture of the platform. In the results section, we show several application examples to illustrate the flexibility and typical use cases of CoVex. Finally, we will discuss opportunities and limitations in using CoVex for COVID-19 research.

CoVex opens up the systems medicine toolbox for the entire infectious disease research community by providing an easy-to-use web tool enriched with novel algorithms. This allows specialists from different fields to bring in expert knowledge to identify the most promising drug targets and drug repurposing candidates for developing effective therapies. We like to stress that the CoVex platform can and will be adopted and extended to allow exploring other viral-host-drug interactomes, e.g. with MERS (Middle East Respiratory Syndrome), Zika, dengue and influenza viruses, thereby increasing preparedness for similar future events.

## Methods

### *Data Integration*

We integrated virus-host interaction data from several sources. We obtained SARS-CoV-2 affinity purification-mass spectrometry (AP-MS) data reported by Gordon *et al.*[10] containing 332 high-confidence virus-host interactions for 27 SARS-CoV-2 proteins[10], as well as SARS-CoV-1 interactions from VirHostNet [4] (24 interactions), and Pfefferle *et al.* [12] (113 interactions existing in our interactome). Human PPIs were obtained from the integrated interactions database (IID) [13] filtered based on experimental validation. The resulting interactome consists of 17,666 proteins connected via 329,215 interactions. Drug-target associations were obtained from ChEMBL (2020-03) [14], DrugBank (v.5.1.5) [15], DrugCentral (2018-08-26) [16], Target Therapeutic Database (2019-07-14) [17], Guide To Pharmacology (2020-01; only approved drugs) [18], PharmGKB (downloaded 2020-04) [19], and BindingDB (2019-08-12) [20]. Where applicable, we considered drugs that have binding affinity values (EC50, IC50, Kd, and Ki) less than 10 μM [21,22]. Only drugs that were mappable to DrugBank IDs and targeting host proteins were included in the network. Drugs currently undergoing clinical trials and mappable to DrugBank IDs (as of April 4th 2020) for the treatment of COVID-19 were collected from ClinicalTrials.gov (www.ClinicalTrials.gov) [23], the EU Clinical Trials Register (www.clinicaltrialsregister.eu) and the International Clinical Trials Registry Platform (www.who.int/ictrp/). In total, we have 6,861 drugs (69 in clinical trials) and 52,860 drug-target associations integrated in our network.

*Systems medicine algorithms for drug repurposing prediction*

The general idea of CoVex is to provide researchers and clinicians with a tool to visually explore druggable molecular mechanisms that drive the interactions between virus and host. To this end, the integrated virus-human-drug interactions form molecular networks that are modeled as graphs with nodes as proteins or drugs, and edges referring to interactions between them. The goal of CoVex is to explore this network while allowing for the exploitation of expert knowledge. Starting with a selected set of (usually) hypothesis-driven seeds (virus proteins, human proteins, or drugs), the goal is to first identify subnetworks connecting these seeds and, subsequently, to identify drug repurposing candidates associated with these mechanisms. A vast number of methods have been reported in the literature for identifying subnetworks [24]. In CoVex, we have integrated several algorithms (including a dedicated, new development) with different underlying paradigms to provide specific exploration options to various particular medical, therapeutic, and research questions and hypotheses. CoVex, thus, allows users to choose among the following approaches in the “advanced analysis” procedures.

**Degree centrality** is the simplest conceivable centrality measure and ranks proteins or drugs interacting with the seeds by their node degree, i.e. the number of interactions. Thus, this algorithm yields subnetworks in which seed-connected proteins and/or drugs are preferentially selected if they interact with many other proteins in the network. The only user-selected parameter is the result size, i.e. how many of the top-ranked proteins or drugs are included. Notably, centrality measures in CoVex can be used for detecting drug targets and for identifying promising drugs.

**Closeness centrality** is a node centrality measure that ranks the nodes in a network based on the lengths of their shortest paths to all other nodes in the network. The rationale behind this algorithm is to preferentially select proteins and/or drugs that are a short distance from all other proteins in the network and are thus of central importance. In CoVex, we use a modified version suggested by Kacprowski *et al.*, where only the shortest paths to a set of selected seed nodes are considered [25]. The only algorithm-specific, user-selected parameter is the result size.

**TrustRank:** TrustRank is conceptually similar to closeness centrality but additionally considers the importance of the seed nodes themselves. In other words, TrustRank ranks nodes in a network based on how well they are connected to a (trusted) set of seed nodes [26]. It is a variant of Google's PageRank algorithm, where "trust" is iteratively propagated from seed nodes to neighboring nodes using the network structure. The node centralities are initialized by assigning uniform probabilities to all seeds and zero probabilities to all non-seed nodes. In CoVex, the TrustRank algorithm can be run starting from a user-defined set of (trusted) seed proteins to obtain a ranked list of proteins in the PPI network that could be prioritized as putative drug targets. Similarly, TrustRank can be executed on the joint protein-drug interactome to identify drug repurposing candidates. User-selected parameters include the result size and the damping factor (range 0 to 1), which controls how fast "trust" is propagated through the network. A small damping factor results in a conservative behavior of the algorithm (nodes close to the seeds receive much higher scores than distant ones), while a large damping factor makes its behavior more explorative.

**Multi-Steiner tree*:*** The Steiner tree problem is a classical combinatorial optimization problem. It aims at finding a subgraph of minimum cost connecting a given set of seed nodes. For CoVex, we have developed a novel method (weighted multi-Steiner tree) which computes approximate weighted multiple Steiner trees and connects them to one subnetwork. The user can select the set of proteins of interest and extract sub-network(s) that connect the selected seed proteins as candidate mechanism(s) involved in COVID-19 progression. In this mechanistic subnetwork(s), we can then extract essential proteins and, thus, the most promising drug targets and repurposable drugs for COVID-19. User-selected parameters include the number of Steiner trees to be merged as well as the tolerance towards accepting more expensive subnetworks (for speeding up the approximation algorithm; details in the Supplementary Material).

**KeyPathwayMiner** is a network enrichment tool that identifies condition-specific subnetworks (key pathways) [27]. In CoVex, we utilize the KeyPathwayMiner web service to extract a maximally connected subnetwork starting from a user-defined set of proteins of interest (seeds). The only user-selected parameter is K, which represents the number of permitted exception nodes, i.e. proteins that were not part of the seed proteins but serve to connect them. Since these proteins act as bridges, these may represent key proteins participating in the dysregulated subnetwork even though they are not directly targeted by the virus and are therefore promising candidates for intervention. In its current implementation exception nodes will only be added if they indeed possess a bridging characteristic and will not be shown otherwise.

**Hub penalty:** Irrespective of the network analysis method used, the extracted solutions have a higher intrinsic probability to contain high-degree nodes (hubs), i.e. proteins that have a large number of interactions. While these proteins are key players in the human interactome, they are not necessarily suitable drug targets as perturbing them might lead to severe unintended side-effects. To mitigate this bias, users can either select an upper bound to filter out high-degree nodes or, alternatively, penalize high degree nodes by incorporating the degree of neighboring nodes as edge weights in the optimization. For the latter, values between 0 and 1 can be selected, where higher values correspond to a higher penalty. Both options are available in advanced analyses for all methods except for degree centrality, because its rationale is to identify hubs, and KeyPathwayMiner, which conceptually does not allow for weighted subnetwork extraction.

*Implementation*

CoVex consists of five components: (i) Data is stored in a PostgreSQL database (v. 12.2). (ii) The backend is implemented using the Django web framework (v. 3.0.5) with Python (v. 3.6) and the Django REST framework (v. 3.11.0) to build the web API. (iii) The network algorithms (except KeyPathwayMiner) are implemented with graph-tool (v. 2.3.1) [28]. (iv) Background task processing is implemented using Redis Queue (RQ, v. 1.3.0) and the in-memory database Redis (v. 3.4.1). Django enqueues the jobs and RQ processes them in the background while Redis functions as a broker between Django and RQ. (v) The frontend is implemented in Angular (v. 9.0.2) and utilizes the JavaScript libraries vis-data (v. 6.5.1) and vis-network (v.7.4.2) for network visualization.

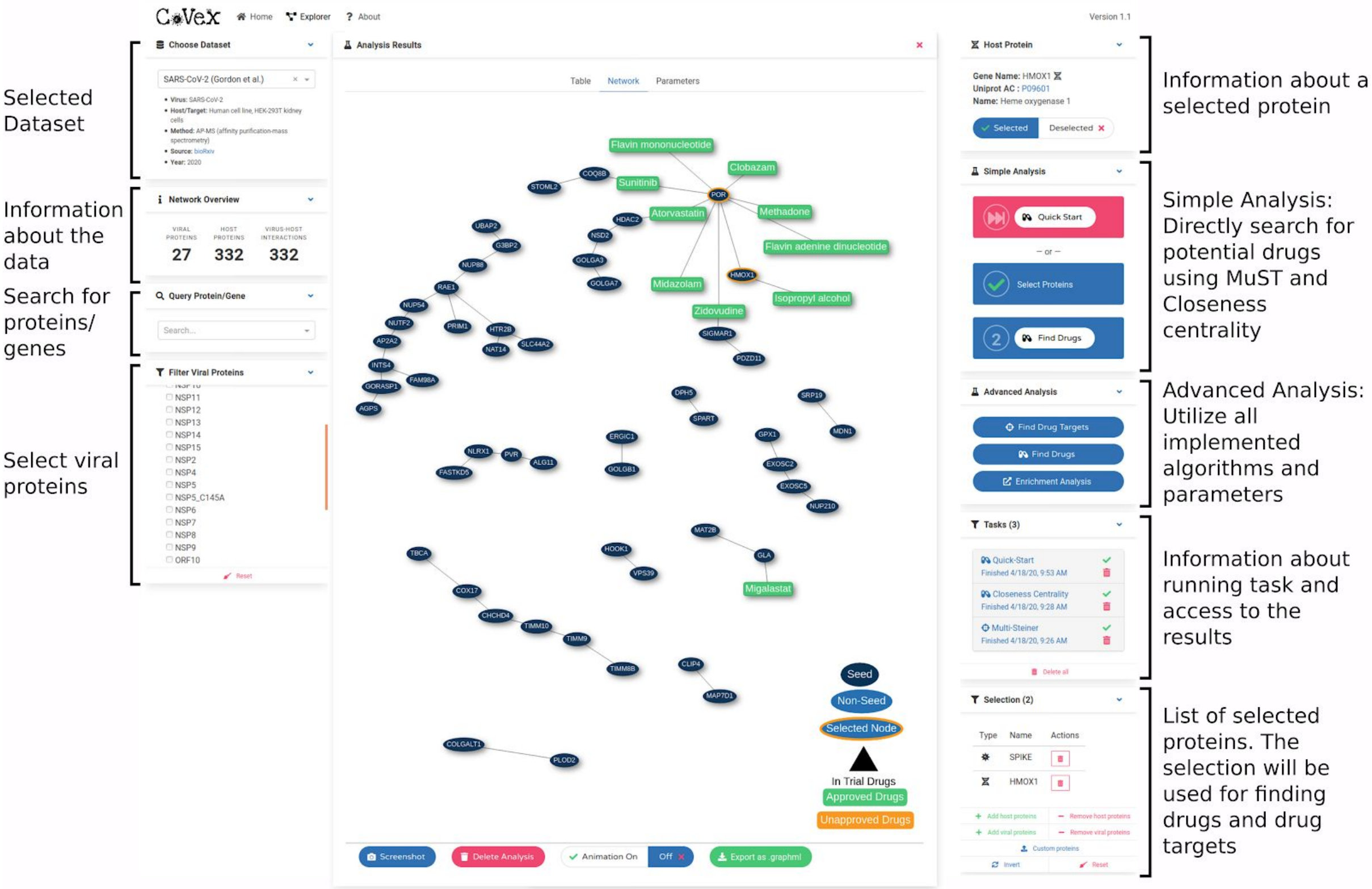

**Figure 2 - The CoVex online platform.** The network view (middle) shows drug candidates (green nodes) that were found using closeness centrality on a set of proteins (blue nodes) which resulted from a multi-Steiner tree computation with all viral proteins as seeds (not shown here). Therefore, drugs targeting these seeds might be able to interrupt the viral life cycle progression.

## Results

The main result is the CoVex platform itself, which renders drug repurposing research systems-medicine-ready. In the following, we first describe how the platform's user interface provides the full feature spectrum of CoVex to clinicians and scientists. Afterwards, we demonstrate the use of CoVex in four different application scenarios starting with four hypotheses and ending with different drug repurposing candidates as well as a short discussion on how to prioritize them.

*The CoVex platform*

Figure 2 shows the CoVex web interface. To find potential drugs, the "Quick Start" analysis will produce a multi-Steiner tree, which considers all viral proteins as seeds and adds a small number of host proteins to connect them. Subsequently, drugs directly targeting these proteins are selected via closeness centrality. After the computation has finished, a click on the corresponding task opens the analysis results, consisting of a table view of drugs and proteins, a visualization of the protein-protein and drug-protein interactions, and a list of parameters used for the analysis. In the "Simple Analysis" panel, users can select seed proteins manually and search for drugs targeting them. In the "Advanced Analysis" panel, users can choose from a list of network medicine algorithms (see Methods and Supplementary Material for details) to discover drug targets or drug repurposing candidates. An enrichment analysis of the identified drug target proteins may be performed with g:Profiler [29].

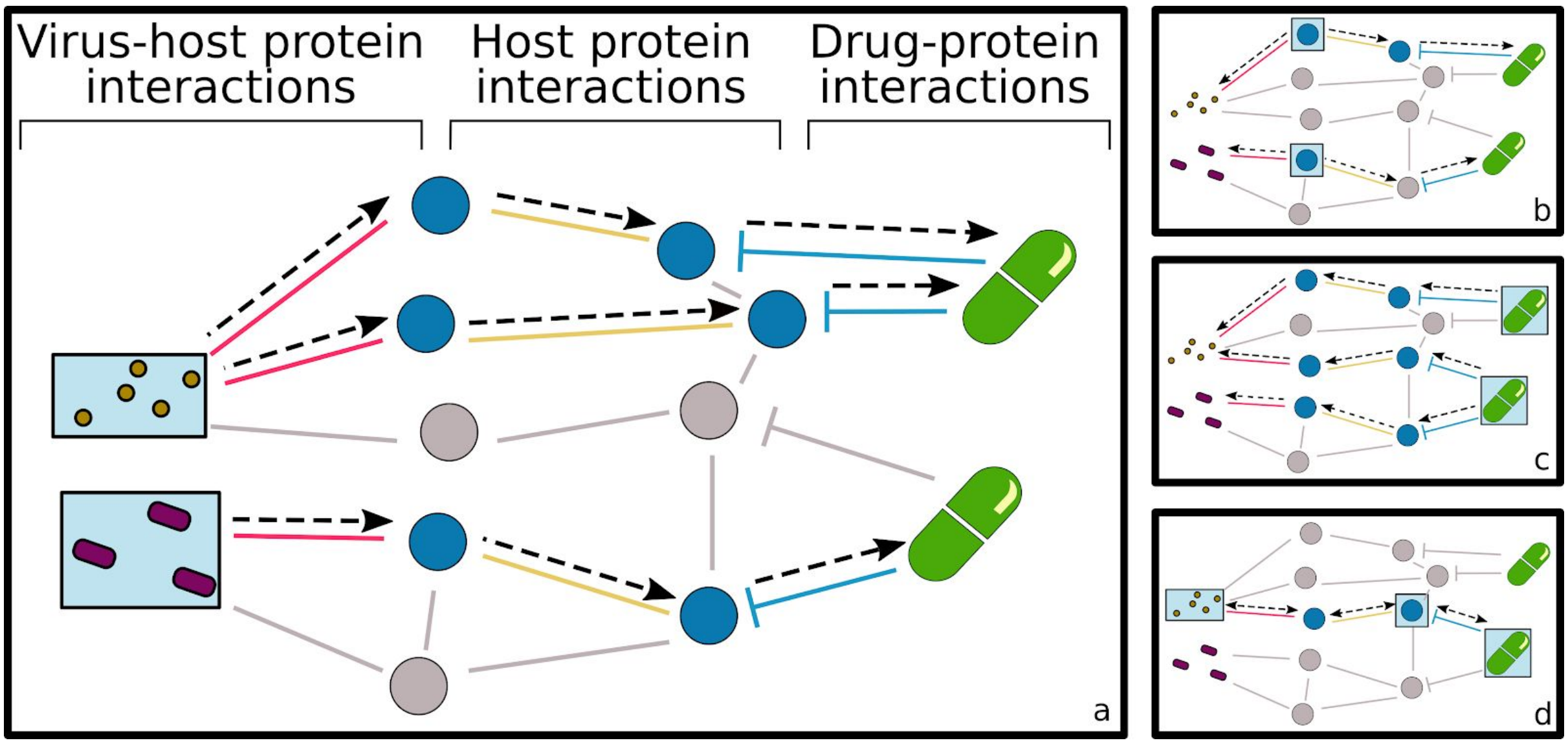

**Figure 3 - CoVex application scenarios.** Depending on the starting hypothesis, dedicated systems medicine algorithms will propagate from selected seeds to connect drugs with viral proteins using host proteins as proxies. Essentially, four different strategies apply: (a) Starting with viral proteins, one can identify drugs targeting host proteins that connect the viral seeds. (b) Starting with a set of proteins of interest as proxies, we identify pathways connecting them to (selected or all) viral proteins. Subsequently, we identify drugs targeting this mechanism. (c) Starting with a set of drugs of interest, one may find pathways to (selected or all) viral proteins extracting a potentially druggable host mechanism. (d) Hypothesis-driven, hybrid approach with seeds in different levels to be connected for druggable mechanism extraction. Boxes with light blue background indicate the typical starting points in the respective application scenario.

*Application scenarios*

The utility of CoVex and its integrated systems medicine approaches is outlined in the following four scenarios. More details on each can be found in the Supplementary Material.

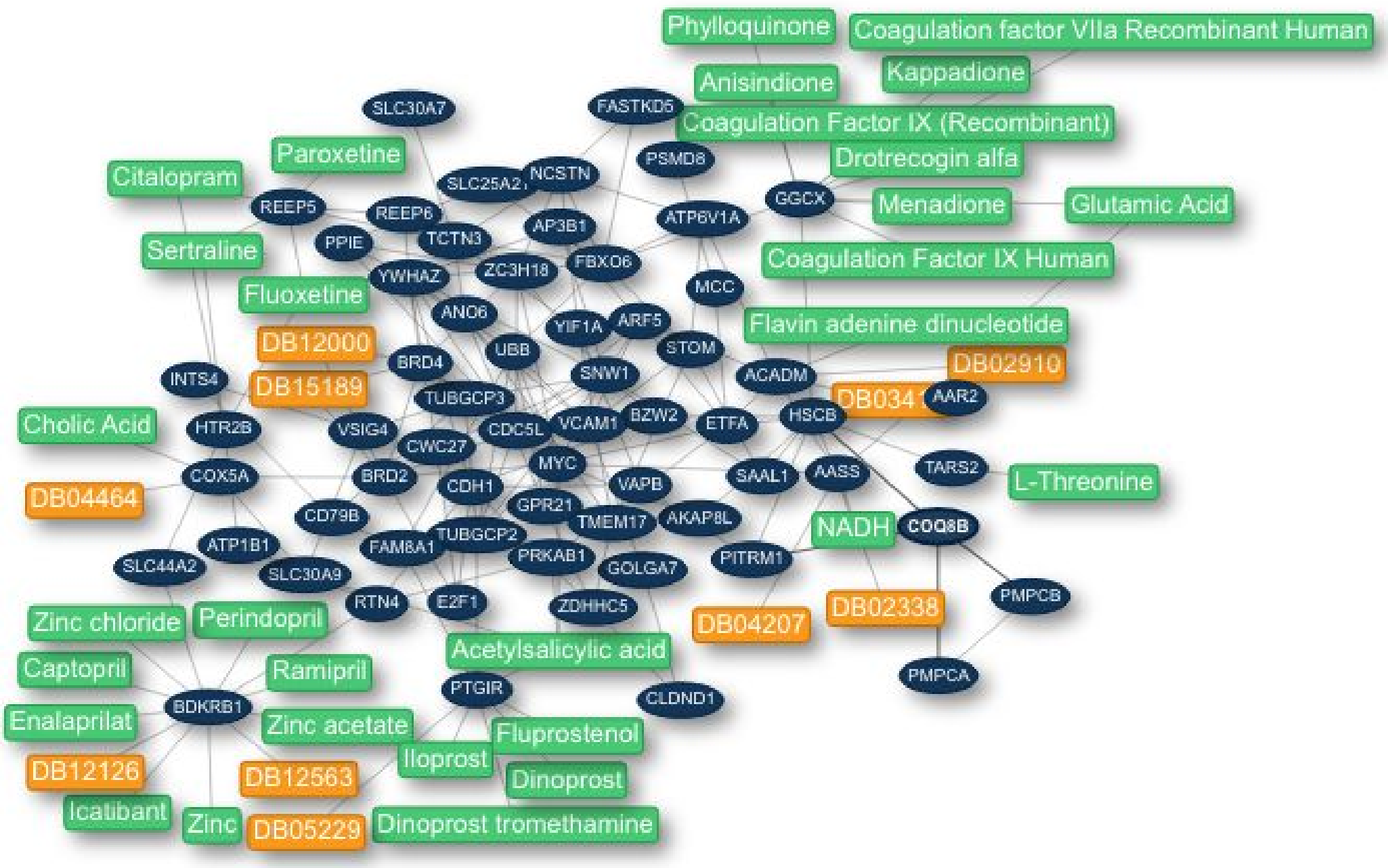

**Figure 4 - Drug-protein-protein interaction network obtained using the viral proteins E, M and Spike with multi-Steiner tree followed by closeness centrality.** Blue nodes are protein targets. Green nodes are approved drugs and orange nodes are non-approved drugs. Lines represent the interactions between proteins and drugs. Note that some ACE inhibitor drugs have been identified, such as Ramipril, Captopril, Perindopril and Enalaprilat targeting the BDKRB1 protein, which are currently being evaluated in clinical trials.

**Scenario a.** Starting from a selection of viral proteins, we use the PPI network to identify the biological mechanism or pathway utilized by the virus. As an example, we consider the viral proteins E, M and Spike, which constitute the external structure of the virus and thus mediate entry into the host cells during the infection process [30,31]. We select the interactors of these viral proteins reported for SARS-CoV-2 and use the multi-Steiner tree algorithm to uncover the biological pathway involved. The resulting network (Figure 4) yields 26 new potential drug targets, including the Bradykinin receptor B1 (BDKRB1). Subsequently, we use closeness centrality to find drugs affecting this pathway. Notably, we identify 6 relevant drugs that target BDKRB1: Ramipril, Captopril, Perindopril and Enalaprilat (approved), which belong to the Angiotensin Converting Enzyme (ACE) inhibitor class [32]; Icatibant, an antagonist of the Bradykinin receptor B2 [33]; and bradykinin, a non-approved drug which is degraded by the ACE [34]. Furthermore, to understand the relationship between BDKRB1 and two proteins known to participate in the entry of the virus (Angiotensin Converting Enzyme 2, ACE2 and Transmembrane protease serine 2, TMPRSS2) [35], we use the “custom proteins” option available in CoVex. We find that Kininogen 1 (KNG1) and Angiotensin (AGT) proteins connect BDKRB1 with ACE2. These 4 proteins are functionally related through the Renin-Angiotensin System, which is targeted by ACE inhibitors (www.wikipathways.org/instance/WP554). In summary, CoVex identifies the protein BDKRB1, which appears to play a role in SARS-CoV-2 host cell entry and can be targeted by several ACE inhibitors widely used in clinical trials to treat COVID-19. It should be noted that the ACE2 protein is not present in the set of seeds used to start the analysis. Nevertheless, CoVex is capable of identifying the pathway and new protein targets functionally related to ACE2 (Figure 4).

**Scenario b.** Starting from both viral proteins and a list of proteins of interest, we can use CoVex to identify a connecting pathway or biological mechanisms that can be targeted by drugs. In this scenario, we are specifically interested in viral proteins that suppress host immunity and the corresponding host immune response pathways. First, we select the viral proteins ORF7a and ORF3a, which are potentially involved in innate immune response and apoptosis as discussed in Gordon *et al.* (2020) [10]. Next, we compile a list of proteins of interest based on the differentially expressed genes (DEGs) from the study by Blanco-Melo *et al.* (2020) [36] lung epithelial cells were infected with the SARS-CoV-2 virus, leading to altered expression of immunity-related genes to combat the viral infection. We consider DEGs known to be associated with the host pathway involving infection with the Herpes simplex virus (HSV), another viral pathogen. These genes include *IFIH1*, *OAS1*, *STAT1*, *DDX58*, *OAS2*, *OAS3*, *IRF7*, *EIF2AK2*, *IFIT1*, and *IRF9*. The selected viral proteins and DEGs (converted to Uniprot ids) were used as seeds for the multi-Steinert tree algorithm to extract a potential immune-related mechanism. As expected, the resulting subnetwork reveals that the viral proteins are close to the DEGs in the host PPI network. Closeness centrality analysis assigned a high rank to Tofacitinib and Ruxolitinib, which are currently being assessed in clinical trials. Tofacitinib and Ruxolitinib exert immunomodulatory effects as Janus kinase (Jak) inhibitors [37,38]. Thus, administration with these drugs may mitigate immune-mediated lung injury and reduce functional deterioration caused by an over-amplified host inflammatory response. This could be especially important in later stages of the disease to prevent an overreaction of the body's immune system and, hence, may further prevent the need for mechanical ventilation in patients suffering from severe COVID-19. Other drugs that target this subnetwork include Masitinib, Erlotinib, and Sorafenib, which could be further examined in downstream analyses. In a similar manner, users may provide a custom list of proteins as seeds to hunt for drugs that can target a putative mechanism of interest.

**Scenario c.** Starting with a set of drugs of interest, we can follow a top-down approach to extract potential host mechanisms and additional drugs targeting the proteins participating in these mechanisms. As an example, we identify 69 drugs currently in clinical trials for COVID-19 and group them based on their Anatomical Therapeutic Chemical (ATC) classification (Suppl. Table S5)[39]. We focus on drugs from the immunostimulants class (L03) and their target proteins as starting seeds. We further select the interactors of the immune-related viral proteins ORF9B, ORF6, ORF3B, and ORF3A[10] as end-point seeds. By applying the multi-Steiner tree algorithm, we discover pathways of interacting proteins that connect the selected drugs (and their target proteins) with the selected viral proteins. Among these connector proteins, we find five genes associated with cytokine signaling in the immune system according to Reactome Pathways CSF2, NRG1, NUP188, PTPN18,

SOCS1)[40]. Notably, CSF2 is enriched in lung, pancreas and immune cells (www.proteinatlas.org/ENSG00000164400-CSF2)[41] and can be inhibited by KB002 (DB05194), which is an investigational drug and an engineered human monoclonal antibody treatment for inflammatory and autoimmune processes[15]. In summary, with CoVex, we found a new drug target which may play a key role in the host immune response during viral infection. We also identified a new drug candidate for COVID-19, as it targets the proteins involved in the pathogenic mechanisms triggered by ORF3A, ORF3B, ORF6, and ORF9B viral proteins.

**Scenario d.** Starting from a hypothesis-driven mixed selection of viral and host proteins as well as drugs, we seek to utilize PPIs to identify a full mechanism or pathway and to suggest additional drug candidates. As an application case, we follow up on a recently published hypothesis by Liu and Abrahams concerning the putative interference of SARS-CoV-2 with the formation of hemoglobin in erythrocytes [42,43]. Essentially, the virus is believed to interfere with heme formation causing symptoms of hypoxia. Liu and Abrahams hypothesize that this would also explain why Chloroquine and Favipiravir are effective drugs, as they may prevent the viral proteins from competing with iron for the porphyrin in hemoglobin (NSP1-16, ORF3a, ORF10, and ORF8 targeted by Chloroquine as well as ORF7a targeted by Favipiravir). Based on this hypothesis (discussed in more detail in the Supplementary Material), we investigate the pathways connecting these viral proteins with the two effective drugs Chloroquine and Favipiravir. To this end, we select two known heme binding host proteins as seeds: Cytochrome b5 reductase, which interacts with the viral protein NSP7, and the viral ORF3a, which binds to Heme oxygenase 1 (HMOX1). Using KeyPathwayMiner for drug target discovery followed by Closeness Centrality for drug discovery, we identify methylene blue in addition to Chloroquine and Deferoxamine, which are both in COVID-19 clinical trials [44,45]. Notably, Methylene blue is approved by the FDA for the treatment of methemoglobinemia, which fits the investigated hypothesis (reduced oxygen-carrying capacity). Also, Deferoxamine is widely used therapeutically as a chelator of ferric ions in disorders of iron overload [46]. However, note that the available scientific evidence for a methemoglobinemia or ferric ion imbalance caused by SARS-CoV-2 is very limited (see Supplementary Material) and that we use this hypothesis solely to illustrate the potential of CoVex' network medicine investigation and hypothesis testing capabilities.

## Discussion

COVID-19 is a threat to our health, our social life, as well as our healthcare and economic systems around the globe. Since the development of safe and effective vaccines is a time-consuming process, the only alternative to mitigate the damage by the SARS-CoV-2 pandemic is to quickly identify agents for the treatment and control of COVID-19 symptoms. Much attention in biomedical and clinical research is, thus, given to the task of identifying therapeutically exploitable drugs. A particular interest lies in drug repurposing, since already approved drugs can go through shortened clinical trials within months rather than years. While a number of promising drug repurposing candidates are currently being tested, the discovery of such candidates is still unstandardized and mostly unstructured. Systems and network medicine offer alternative approaches, where the process of drug target discovery is driven by computational data mining methods utilizing molecular interaction networks. As recently demonstrated by Gysi *et al.* for SARS-CoV-2, this data-driven process can produce a list of promising drug candidates targeting host proteins in close proximity and mechanistically related to virus-interacting proteins [11]. Here, we seek to make this network medicine approach widely available to the community.

With CoVex, we present an interactive and user-friendly web platform that integrates published data of SARS-CoV-1 as well as recent data about virus-host interactions in SARS-CoV-2 [10] with the human interactome and several drug-target interaction databases. CoVex allows users to mine the integrated virus-host-drug interactome for promising drug targets and drug repurposing candidates with only a few mouse clicks. Through features such as interactive seed protein selection, filtering and upload of own lists of proteins or drugs of interest, CoVex covers diverse application scenarios ranging from data-driven, hypothesis-free drug target discovery to expert-guided analyses with a clear underlying hypothesis about virus biology. To address the diversity of research questions adequately, CoVex implements several state-of-the-art graph analysis methods. These were specifically tailored to be employed in a network medicine context and include a weighted version of TrustRank as well as a novel multi-Steiner tree method (Supplementary Material).

While CoVex is a powerful tool for SARS-CoV-1 and -2 research, results uncovered with our platform have to be considered with caution. We stress that the suggested drug candidates need to be properly vetted by clinical experts and tested following established procedures and clinical trials. Current data about virus-host interactions in SARS-CoV-2 is still preliminary and incomplete. For instance, important proteins such as the ACE2 receptor, a known entrypoint for the virus [35], is missing in the SARS-CoV-2 dataset by Gordon *et al.* [10]. Moreover, we included only drugs that are reported in databases about clinical trials or in the literature if they have a valid entry in DrugBank, possibly excluding some of the drugs currently being investigated. Further, we do not differentiate between different sources of drug target interactions. The strength of experimental evidence may vary depending on the experimental assay that was used or the type of annotation from the source database, e.g. clinical annotations and variant annotations from PharmGKB which can be interpreted as indirect drug-protein associations. It should also be noted that we do not list drugs that target viral proteins directly, as the goal of CoVex is to unravel novel drug targets further downstream in the human interactome.

We acknowledge that the choice of algorithm and its associated parameters is non-trivial, forcing users to engage in time-consuming explorative analysis. To make this easier, we allow users to queue multiple tasks, which are executed in parallel. As our experience with this platform grows, we also plan to develop guidelines that allow users to choose an appropriate method for a particular research question. We further plan to integrate new data about virus-host interactions and ongoing clinical trials in corona viruses as it becomes available.

For the future, we also plan to extend the CoVex network medicine platform to other viruses in which new drug targets and drug repurposing candidates are urgently sought, including MERS, Zika, influenza and dengue.

## Conclusions

We present CoVex, the first web-based platform for the interactive exploration and network-based analysis of virus-host interactions, aimed towards drug repurposing for the treatment of COVID-19. CoVex can be easily updated to accommodate the fast-paced data generation in the battle against the global pandemic. CoVex is expected to speed up the discovery of potential therapeutics for COVID-19.

## Acknowledgements

TK and SS are grateful for financial support from H2020 project RepoTrial (nr. 777111). JM, JS, NKW, and RN received funding from H2020 project FeatureCloud (nr. 826078). JB, TK, and ML are grateful for financial support from BMBF grant Sys_CARE (nr. 01ZX1908A) of the Federal German Ministry of Research and Education. JB's BMBF grant SyMBoD (nr. 01ZX1910A) also financed parts of this project. MO is funded by the Collaborative Research Center SFB924 of the German Research Foundation. JB was partially funded by his VILLUM Young Investigator Grant nr. 13154. Contributions by JKP and TDR are funded by the Bavarian State Ministry of Science and the Arts in the framework of the Centre Digitisation.Bavaria (ZD.B, grant LipiTUM). MSA is grateful for a PhD fellowship funding from CONACYT (CVU659273) and the German Academic Exchange Service, DAAD (ref. 91693321).

**Supplementary Material - Exploring the SARS-CoV-2 virus-host-drug interactome for drug repurposing**

Here, we present the new algorithmic techniques we developed to analyze the virus-host-drug interactome by starting from a set of seed nodes, avoiding hubs. Moreover, we give additional, detailed information about the application cases described in our CoVex paper.

**New models and algorithms for seeded, weighted network analysis with hub-penalties**

**Multi-Steiner trees:** The Steiner tree problem is a classical combinatorial optimization problem. Given an undirected graph $G = (V, E)$ with non-negative edge costs $c$ and a set of seed nodes $S \subseteq V$, it asks to compute a minimum-cost Steiner tree for $S$, where a Steiner tree for $S$ is a tree in $G$ that contains all nodes $v \in S$. Exactly solving this problem is NP-hard, but several efficient approximation algorithms exist.

In CoVex, we compute (several) Steiner trees to connect selected host or viral proteins and extract a potentially druggable mechanism. For this, we developed a customized version of the 2-approximation algorithm by Kou et al. [1,2]. There are two main differences: First, we do not only compute one but return the union of $K$ Steiner trees, where $K$ is a parameter that can be set by the user. Computing several Steiner trees increases the stability of the extracted mechanism. This is important, because the networks used in CoVex are very dense, which implies that solutions to the Steiner tree problem are usually non-unique. Which solution is returned is, hence, a matter of chance, if only one Steiner tree is computed. Returning the union of $K$ Steiner trees mitigates this problem. Second, we employ special parameterized edge costs $c_{\lambda}$ with a hub-penalty $\lambda \in [0, 1]$ that can be specified by the user (detailed below). If $\lambda$ is set to a value close to 1, the returned Steiner trees avoid hub-nodes. This is important, because targeting proteins which are hub-nodes in the human protein-protein interaction (PPI) network potentially results in many undesired side effects.

The multi-Steiner tree algorithm employed in CoVex is implemented as follows: In a first step, we use the algorithm by Kou et al. to compute the first Steiner tree $T$. Moreover, we run a depth-first search to find all bridges in the graph, where a bridge is an edge whose deletion results in the graph being disconnected. Let $L$ be the list of edges in $T$, $C$ be the cost of $T$, and $\tau$ be a user defined tolerance that specifies by how much the costs of the subsequent trees may exceed $C$. Furthermore, let $k$ be the number of already discovered trees (initialized to 1) and $U$ be the set of returned nodes (initialized to the nodes contained in $T$). We iterate the following steps until $k = K$ or $L$ is empty. Subsequently, we return the subgraph induced by $U$.

1. Pop edge $e$ from $L$.
2. If $e$ is a bridge, go to step 1.
3. Temporarily delete $e$ from $G$.

4. Run the algorithm by Kou et al. to compute the next candidate tree $T'$.
5. If the cost of $T'$ does not exceed $C \cdot \frac{100+\tau}{100}$, add the nodes of $T'$ to $U$ and increment $k$.
6. Remove all edges from $L$ which are not contained in $T'$.
7. Reinsert $e$ into $G$.

**Weighted TrustRank**: TrustRank is a variant of Google's pagerank algorithm, where "trust" is iteratively propagated through the network starting from an initial set of (trusted) seed nodes [3]. At termination, each node in the network receives a score which is high if the node is easily reachable from a subset of seed nodes, which themselves assume central positions in the overall topology of the network.

In CoVex, TrustRank can be used for two purposes. First, TrustRank can be used to rank drugs targeting a previously selected or computed set of host proteins. Second, it can be employed to discover potentially druggable host proteins which are relevant for a given set of viral or host seed proteins. Note that, unlike the result returned by the multi-Steiner algorithm, the top ranked nodes returned by TrustRank are not guaranteed to be connected. In practice, however, many of the returned nodes often form one large connected component. Thus, TrustRank allows to extract pathways in situations where it is not clear *a priori* that all seed nodes are involved in one mechanism.

In CoVex, we use a customized version of the TrustRank algorithm which allows the user to avoid hubs, if required. This is achieved by defining parameterized edge diameters $d_\lambda(e) = 1/c_\lambda(e)$ (the definition of $c_\lambda$ is given below in section "Parameterized edge costs with hub-penalty"). If the user sets the hub-penalty $\lambda$ to a value close to 1, the diameters of the edges which are incident with hub-nodes are small. Therefore, "trust" flows less easily to the hubs, which implies that they obtain a lower score. Further parameters are the result size, which specifies how many top ranked nodes should be displayed in the result, and the damping factor $d \in [0, 1]$. The damping factor $d$ controls how easily "trust" can flow to nodes which are far away from the seeds. The larger $d$, the higher scores distant nodes receive.

**Weighted, seeded closeness centrality:** Closeness centrality is a simple centrality measure which ranks the nodes in a network based on the average distance of the shortest paths to all other nodes. Kacprowski *et al.* suggested a version of this centrality measure where only the distances to a selected set of seed nodes are taken into consideration [4].

In CoVex, we use a modified version of this approach which uses the parameterized edge costs $c_\lambda$ instead of uniform costs. This customized version of seeded closeness centrality hence assigns lower scores to hubs, if requested by the user. Like TrustRank, closeness centrality can be used both for extracting drug targets and for ranking drugs.

**Parameterized edge costs with hub-penalty:** For each edge $(u, v) \in E$ in the network $G$, the parameterized costs employed by the three algorithms presented above are defined as $c_{\lambda}(u,v) = (1-\lambda)\cdot avdeg(G) + \lambda \cdot \frac{deg(u)+deg(v)}{2}$, where $deg(u)$ and $deg(v)$ are the degrees (i.e., number of links) of the nodes $u$ and $v$, $avdeg(G)$ is the mean degree of all nodes contained in $G$, and $\lambda \in [0,1]$ is a hub-penalty, which can be specified by the user. Note that for $\lambda = 0$, all edge costs equal $avdeg(G)$ (no hub-penalty), while for $\lambda = 1$, the cost of each edge is a direct function of the degrees of its incident nodes (maximal hub-penalty). By setting $\lambda$ to a value between 0 and 1, the user can balance between these two extremes.

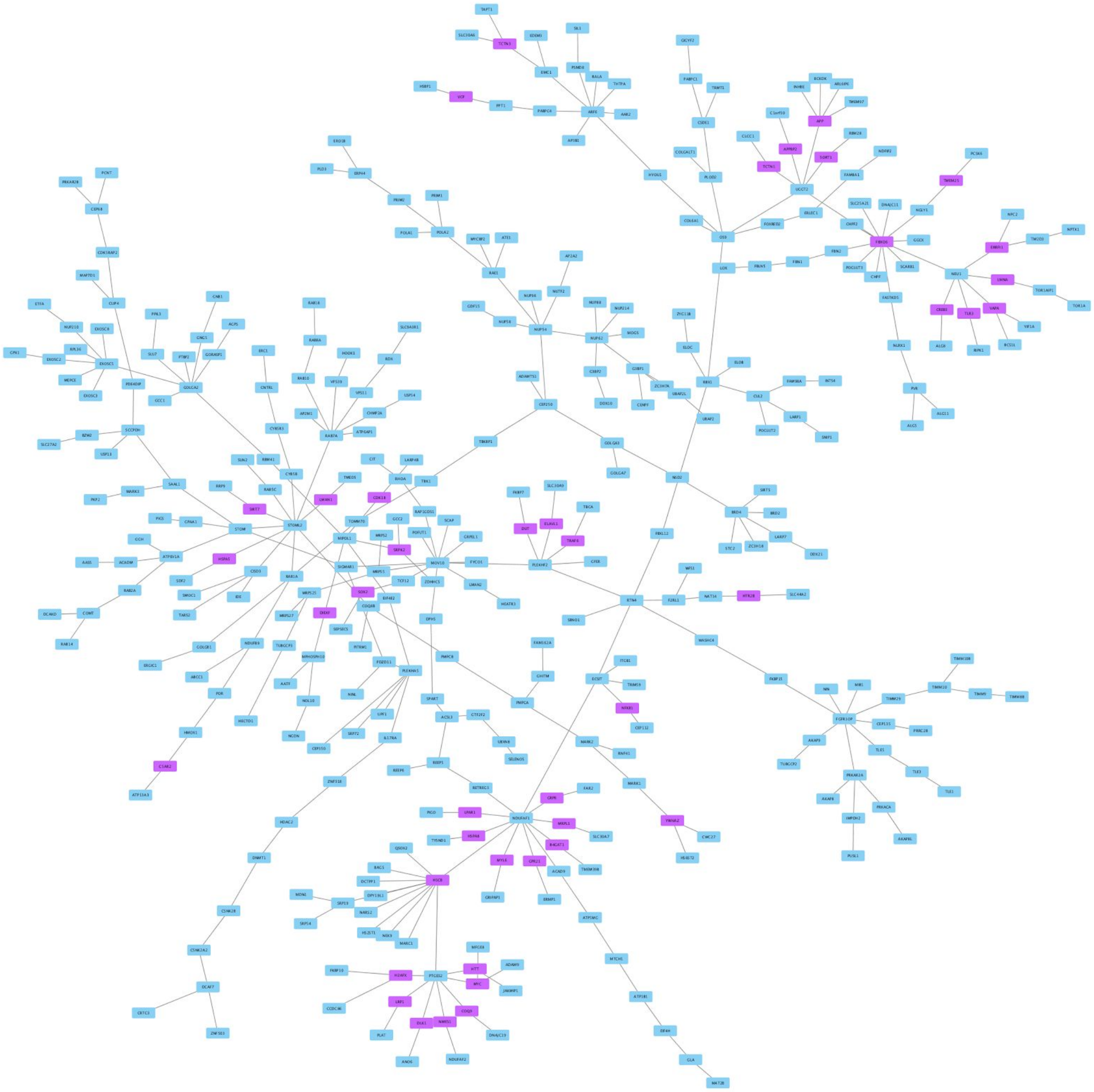

**Figure S1 - The multi-Steiner tree connecting all host proteins** interacting with the virus (blue nodes are seeds) visualized in Cytoscape [5] using the GraphML export feature of CoVex. The purple nodes are the connectors building a tree structured subnetwork of the human interactome.

### Why the host protein interactions matter for drug repurposing

247 out of 332 host proteins interacting with the virus (interactors) reported in Gordon *et al.* [6] for SARS-CoV-2 form a large connected component, indicating that most of the direct targets of the virus are in a close proximity in the human interactome. 12 drugs currently in clinical trials (Dexamethasone, Colchicine, Pravastatin, Ribavirin, Ruxolitinib, Bromhexine, Oseltamivir, Noscapine, Ascorbic acid, Tofacitinib, Artenimol, Suramin) target the virus interactors directly. Figure S1 shows a Steiner tree of minimum cost connecting all the interactors. It consists of 44 connector proteins in addition to 332 interactors. The Steiner tree enables us to find new drug target candidates. Importantly, six of the 69 drugs currently in clinical trials (Table S5) target exclusively the connector proteins revealed by our analysis, namely Glycyrrhizic acid, Synthetic Conjugated Estrogens B, Leflunomide, Chloroquine, Deferoxamine, and Thalidomide.

**The four application scenarios**

**Application scenario a -** Starting from a selection of viral proteins, we seek to use the human interactome to identify biological mechanisms or pathways utilized by the virus during infection. As an example, we are interested in the viral proteins E, M and Spike, which constitute the external structure of the virus and thus participate in the entry into host cells [7,8].

First, we select all host proteins interacting with the viral proteins E, M and Spike from the SARS-CoV-2 dataset. We then use the multi-Steiner Tree algorithm with the parameters shown in Table S1 to uncover the biological pathway involved. The resulting network allows the identification of 26 new potential drug targets, including the Bradykinin receptor B1 (BDKRB1).

Next, we use closeness centrality with the parameters shown in Table S1 to find drugs affecting this pathway. We identify a total of 30 approved and 10 non-approved drugs (Figure S2). Notably, we find 6 relevant drugs that target BDKRB1: Ramipril, Captopril, Perindopril and Enalaprilat (approved), which belong to the Angiotensin Converting Enzyme (ACE) inhibitor class [9]. Icatibant is an antagonist of the Bradykinin receptor B2 [10] and bradykinin is a non-approved drug which is degraded by the ACE [11].

Finally, to understand the relationship between BDKRB1 and Angiotensin Converting Enzyme 2 (ACE2) as well as Transmembrane protease serine 2 (TMPRSS2), two proteins known to be involved during virus entry [12], we use the “custom proteins” option available in CoVex and utilize the multi-Steiner tree algorithm with the same parameters as in Table S1. We find that the Kininogen 1 (KNG1) and Angiotensin (AGT) proteins connect BDKRB1 with ACE2 (Figure S3). These 4 proteins are functionally related through the Renin-Angiotensin System, which is targeted by ACE inhibitors (https://www.wikipathways.org/instance/WP554).

In summary, CoVex identifies the protein BDKRB1, which participates in the pathway affected by SARS-CoV-2 and can be targeted by several ACE inhibitors, which are widely used in clinical trials to treat COVID-19. It should be noted that the ACE2 protein is not present in the set of seeds used to start the analysis; however, CoVex is capable of identifying the pathway and new protein targets functionally related to ACE2, which can be targeted by ACE inhibitors as well. In this case, CoVex allows the identification of the mechanism behind the drugs currently considered for treating currently considered for treating COVID-19.

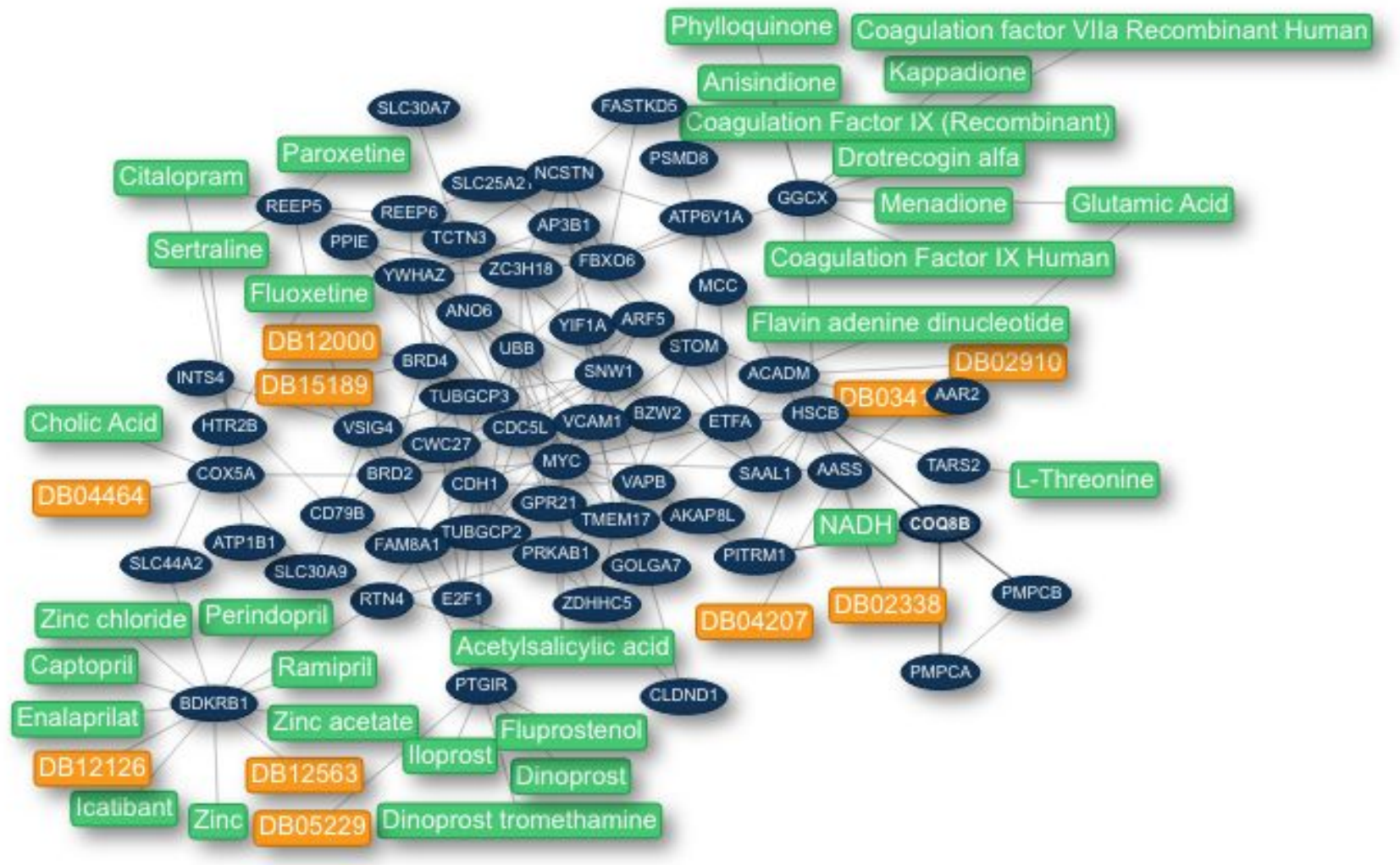

**Figure S2 - Network obtained from interactors of baits E, M and Spike with multi-Steiner tree followed by closeness centrality.** Blue nodes are protein targets, green nodes are approved drugs and orange nodes are non-approved drugs. Lines represent the interactions between the proteins and drugs. ACE inhibitor drugs are identified, such as Ramipril, Captopril, Perindopril and Enalaprilat targeting the BDKRB1 protein, which are currently being evaluated in clinical trials in COVID-19 patients. Note that we also included this Figure in the main document (Figure 4) to showcase how CoVex results look like but kept it here as well for increased readability.

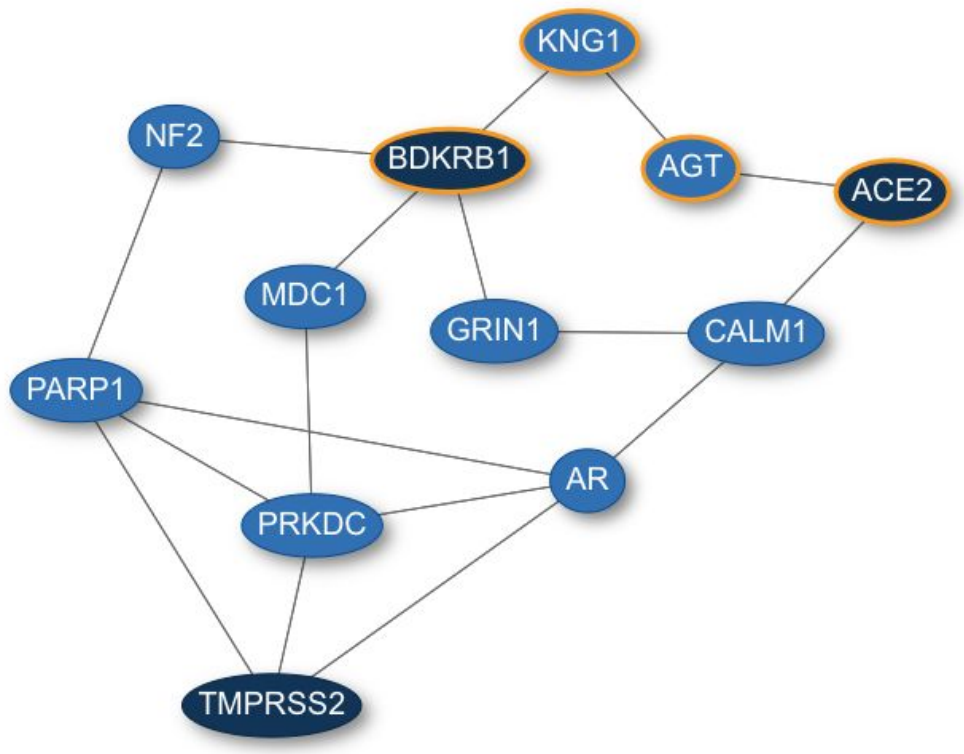

**Figure S3 - The subnetwork obtained with the multi-Steiner tree algorithm shows the connections between ACE2, TMPRSS2 and BDKRB1.** Dark blue nodes represent the seed proteins and light blue nodes represent the connector proteins identified.

**Table S1 - Algorithms and parameters used in application scenario a**

| | Algorithm |
|---|---|

| | multi-Steiner tree | Closeness centrality |
|---|---|---|
| Number of trees | 10 | N.A. |
| Tolerance | 0 | N.A. |
| Result size | N.A. | 40 |
| Maximum degree | DISABLED | DISABLED |
| Hub penalty | 0 | 1 |
| Include non-seed viral proteins | FALSE | N.A. |
| Include indirect drugs | N.A. | FALSE |
| Include non-approved drugs | N.A. | TRUE |

**Application scenario b -** Candidate drugs for the treatment of COVID-19 can be identified starting from a user-defined set of seeds comprising differentially expressed genes (DEGs) and viral proteins. Such a custom list can be obtained from other datasets, such as experiments and/or literature. One possible strategy is to use proteins known to be associated with a specific biological process, such as SARS-CoV-2 proteins involved in viral pathogenesis and host proteins that participate in the corresponding host immune response to infection. This way, we use the host PPI network to connect the viral proteins to the DEGs and obtain a potential mechanism that can be targeted using repurposable drugs. To obtain a custom list of DEGs, raw counts from the gene expression data of SARS-CoV-2-infected lung epithelial A549 cells relative to mock-treated cells were obtained from Blanco-Melo *et al.* [13] (GEO accession GSE147507). Differential expression analysis using the edgeR (v.3.26.8) package (fold change ≥2, adjusted p-value <0.05) was performed to obtain a list of DEGs. To identify host cell pathways enriched in response to SARS-CoV-2 infection, KEGG enrichment was performed using the gseapy (v.0.9.15) package. Enriched KEGG terms included pathways that are known to be involved in immune response to pathogens, such as “Influenza A”, “Herpes simplex infection”, “Measles”, and “Hepatitis C”.

In this example, we use the CoVex platform to select viral proteins involved in innate immune response and apoptosis, namely, ORF7a and ORF3a, as indicated in Gordon *et al.* [6]. Next, using the “Custom proteins” option, we upload the Uniprot IDs of the DEGs that participate in the enriched pathway Herpes simplex infection, which is involved in response to infection with Herpes simplex virus (HSV), another viral pathogen. The enriched DEGs include *IFIH1*, *OAS1*, *STAT1*, *DDX58*, *OAS2*, *OAS3*, *IRF7*, *EIF2AK2*, *IFIT1*, and *IRF9*. We then use both the viral proteins and DEGs as seeds for the multi-Steiner tree algorithm to extract a subnetwork that is relevant to our pathway of interest, shown in Figure S4. Next, we use closeness centrality on the resulting subnetwork to obtain drugs. The parameters used to run the multi-Steiner tree and closeness centrality algorithms can be found in Table S2. Top-ranking drugs included Tofacitinib and Ruxolitinib, which are currently being assessed in clinical trials for the treatment of COVID-19 (Figure S5). Tofacitinib and Ruxolitinib are both known to inhibit Janus kinase (Jak), which promote cytokine signaling [14,15]. Thus, administration with these drugs can mitigate immune-mediated lung injury and prevent functional functional deterioration in COVID-19 patients caused by an over-amplified host immune response. As shown in Figure S5, other drugs that target this subnetwork include Masitinib, Erlotinib, and Sorafenib, which could be further examined in downstream analyses. In a similar manner, users can provide a custom list of proteins to retrieve drugs that can target their mechanism of interest, followed by careful examination of the results.

**Table S2 - Algorithms and parameters used in application scenario b**

| | Algorithm | |
|---|---|---|
| | multi-Steiner tree | Closeness centrality |
| Number of trees | 10 | N.A. |
| Tolerance | 0 | N.A. |
| Result size | N.A. | 20 |
| Maximum degree | DISABLED | DISABLED |
| Hub penalty | 0 | 0 |
| Include non-seed viral proteins | TRUE | N.A. |
| Include indirect drugs | N.A. | FALSE |
| Include non-approved drugs | N.A. | FALSE |

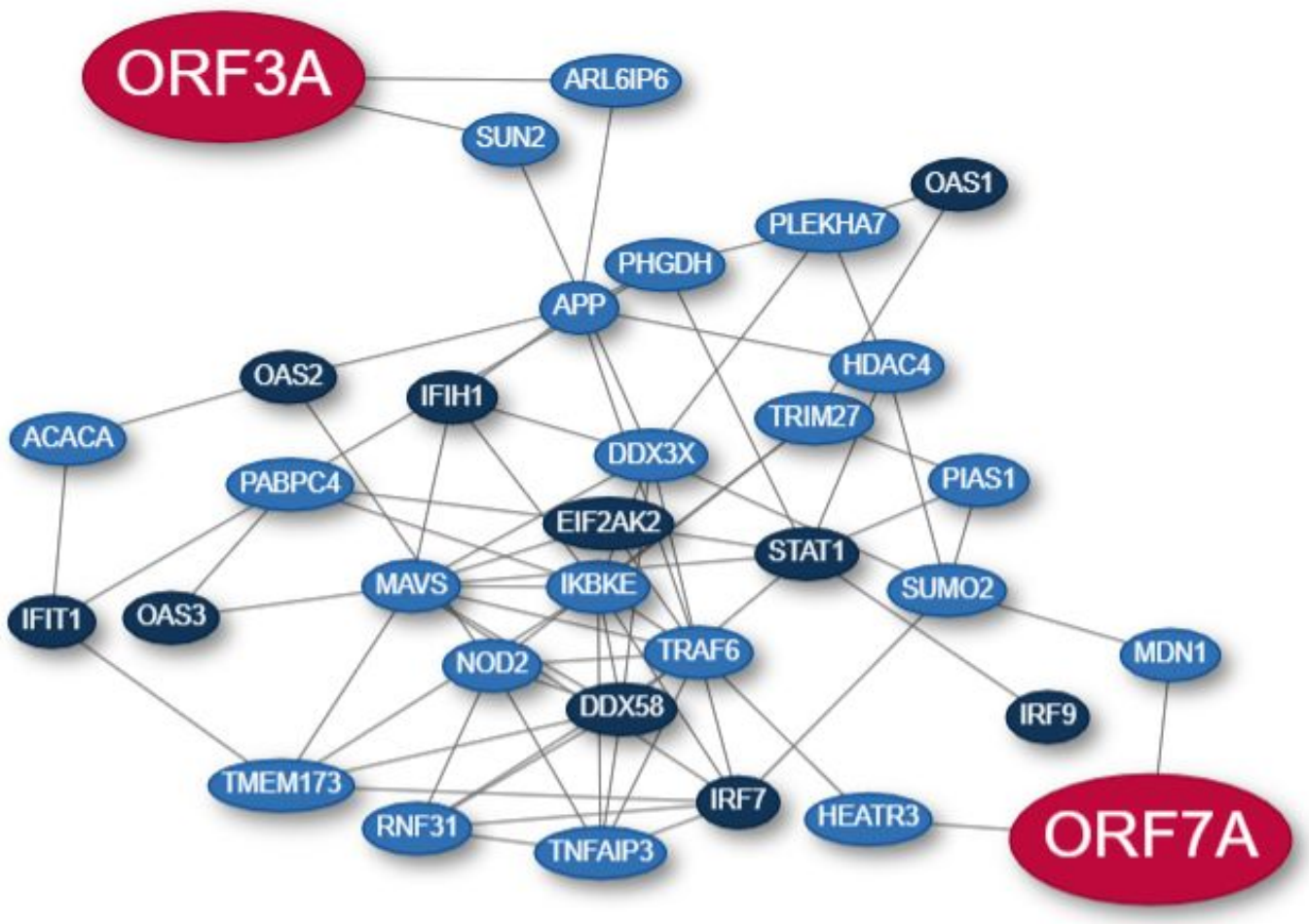

**Figure S4 - The subnetwork showing the connections between the viral proteins (ORF7A, ORF3A) and proteins involved in Herpes simplex infection.** Dark blue nodes indicate proteins involved in Herpes simplex infection, while bright blue nodes indicate the connector nodes found by the multi-Steiner tree algorithm. The resulting subnetwork comprises potential drug targets to suppress the host immune response.

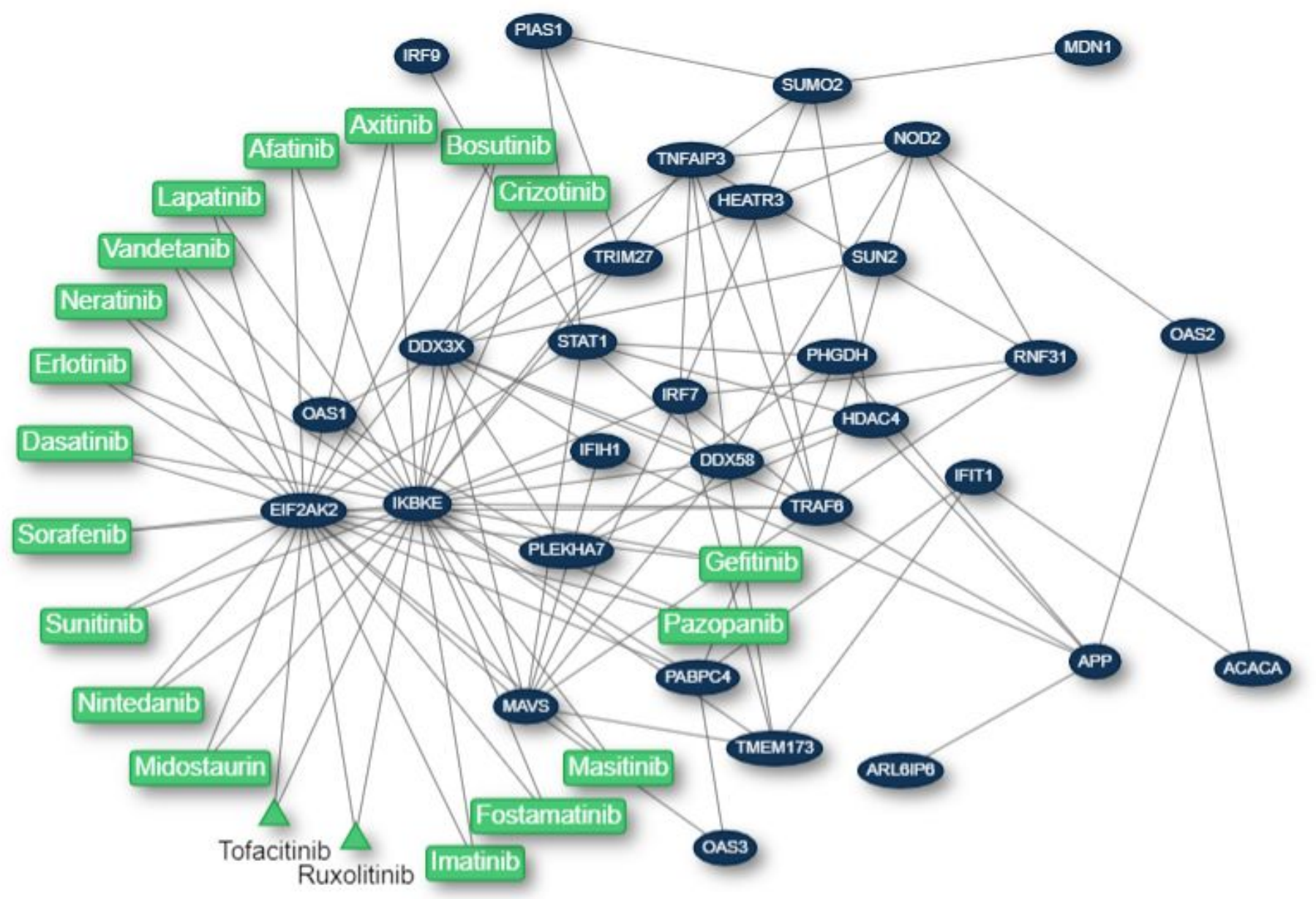

**Figure S5 - Closeness centrality recovered drugs**, such as Tofacitinib and Ruxolitinib, which are currently being evaluated in clinical trials in COVID-19 patients.

**Application scenario c -** Another approach to find candidate drugs to combat COVID-19 is to connect the targets of promising drugs which are already in clinical trials to the viral proteins. We can identify the candidate mechanisms by extracting the sub-network(s) connecting these two ends with a minimum number of intermediate connector proteins. This step can be done by using the multi-Steiner tree algorithm integrated in our CoVex platform. We can then seek for the drugs targeting the identified connector proteins utilizing the closeness centrality algorithm from the “Find Drugs” function.

As an example, we start with the drugs classified as immunostimulants (Sargramostim, Peginterferon beta-1a, and Peginterferon alfa-2a)[16], and then use the multi-Steiner tree algorithm from the “Find Drug Targets” function to connect their targets (IL3RA, CSF2RB, SDC2, PRG2, IFNAR2, IFNAR1, and CSF2RA) to the host interactor genes of viral proteins ORF9B, ORF6, ORF3B, and ORF3A (ALG5, ARL6IP6, BAG5, CHMP2A, CLCC1, CSDE1, DCTPP1, DPH5, HEATR3, HMOX1, MARK1, MARK2, MARK3, MDN1, MTCH1, NUP98, PTBP2, RAE1, SLC9A3R1, STOML2, SUN2, TOMM70, TRIM59, VPS11, VPS39), see Figure S6. Among the identified connector proteins, we find five genes (CSF2, NRG1, NUP188, PTPN18, SOCS1) associated with cytokine signaling in the immune system according to Reactome Pathways [17]. From this list, notably, CSF2 is enriched in lung and immune cells (www.proteinatlas.org/ENSG00000164400-CSF2) [18]. This gene can be inhibited by the investigational drug KB002 (DB05194) (Figure S7), which is an engineered human monoclonal antibody-based treatment for inflammatory and autoimmune processes [19].

**Table S3 - Algorithms and parameters used in application scenario c**

| | Algorithm | |
|---|---|---|
| | multi-Steiner tree | Closeness centrality |
| Number of trees | 10 | N.A. |
| Tolerance | 0 | N.A. |
| Result size | N.A. | 50 |
| Maximum degree | DISABLED | DISABLED |
| Hub penalty | 1 | 0 |
| Include non-seed viral proteins | FALSE | N.A. |
| Include indirect drugs | N.A. | FALSE |
| Include non-approved drugs | N.A. | TRUE |

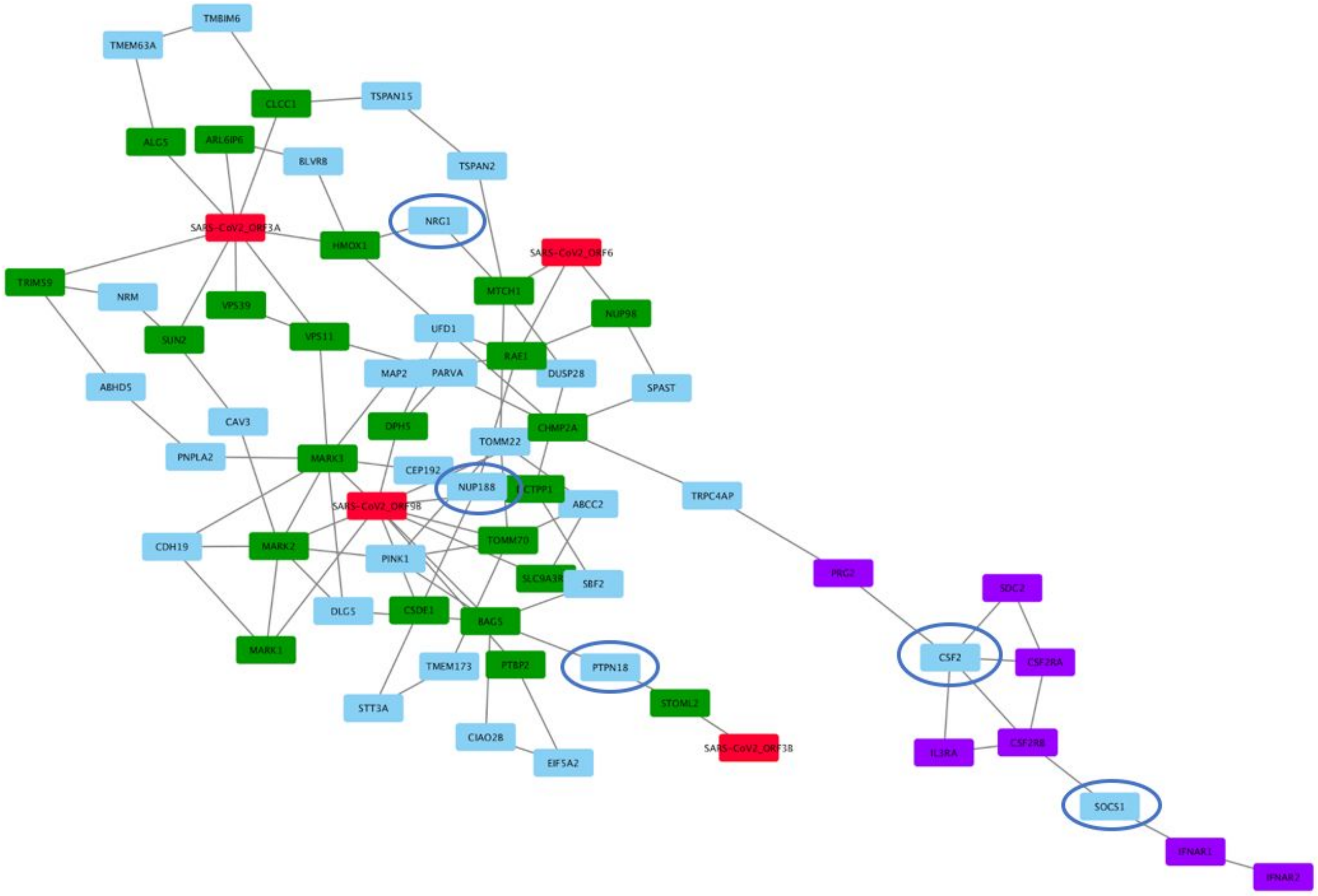

**Figure S6 - Cytoscape illustration of the multi-Steiner tree** analysis performed with CoVex after using the export to GraphML feature, with targets of Sargramostim, Peginterferon beta-1a, and Peginterferon alfa-2a (purple) and the host interactor genes of viral proteins ORF9B, ORF6, ORF3B, and ORF3A (green) as seeds. Connector proteins (blue). Viral proteins (red). Connector genes associated with cytokine signaling in the immune system (circled in blue).

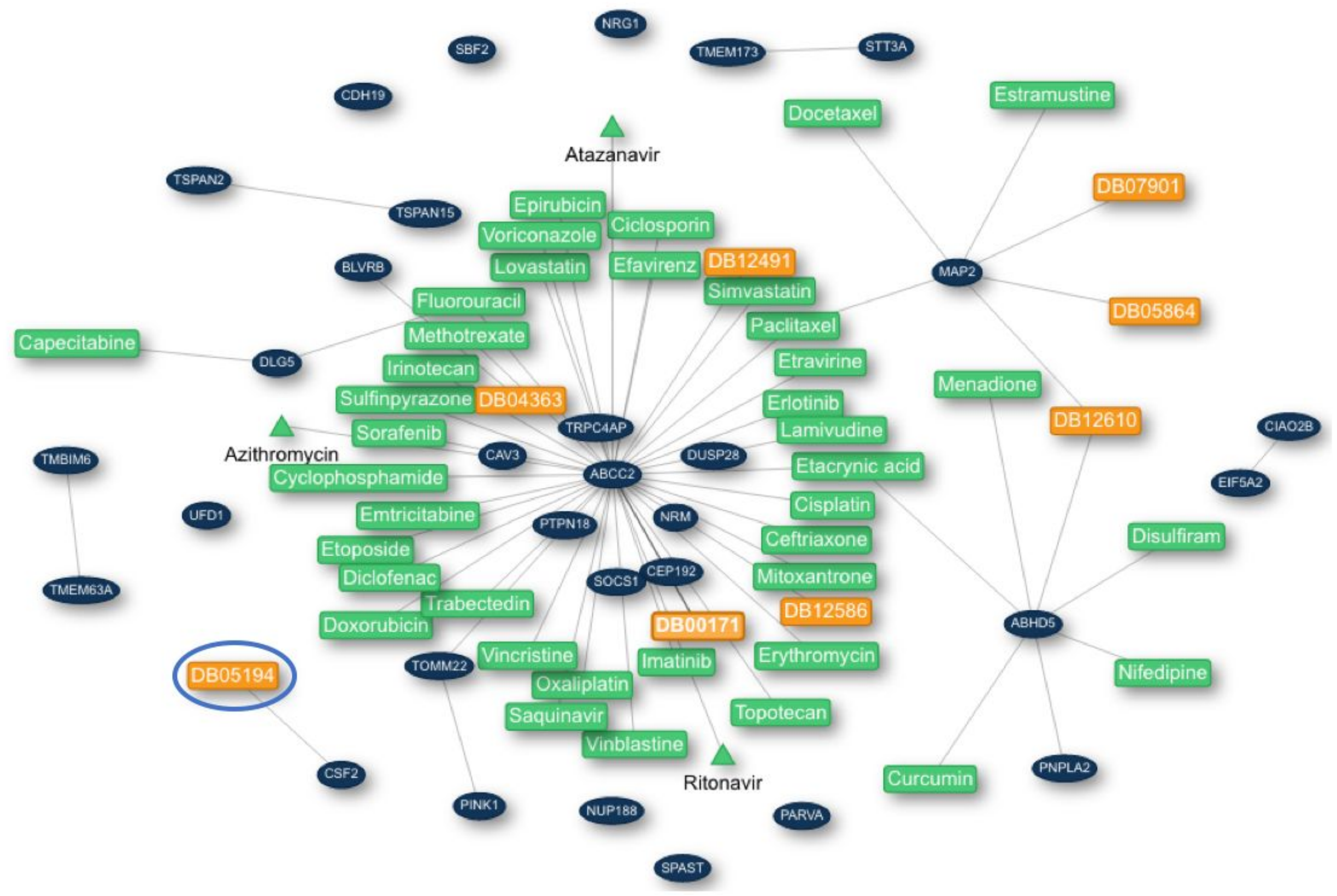

**Figure S7 - The closeness centrality algorithms ranks the drugs** targeting the connector nodes from Figure S6. The protein encoded by CSF2 is inhibited by drug KB002 (DB05194).

**Application scenario d -** Some hypotheses might require a hybrid seeding approach, where we are required to start from a hypothesis-driven mixed selection of viral and host proteins as well as drugs to explore protein-protein interactions to identify a mechanism and suggest additional drugs.

Here, we follow a recently published hypothesis concerning the interference of the SARS-CoV-2 with the formation of hemoglobin in erythrocytes [22,23]. Essentially, we follow the idea that the virus could interfere with porphyrin, which is a substrate, together with iron ($Fe^{2+}$) ions, during the synthesis of the heme prosthetic group in hemoglobin. The viral proteins thus hinder the interaction of iron with porphyrin, as the viral proteins are hypothesized to compete with iron for the porphyrin and thus to inhibit heme group synthesis, leading to hypoxia symptoms [24]. Liu and Abrahams suggest that this might explain why Chloroquine and Favipiravir are effective drugs, as they may prevent the virus from competing with iron for the porphyrin: Chloroquine by interfering with the viral proteins NSP1-16, ORF3a, and ORF10 which may bind the porphyrin to prevent heme synthesis, and also by inhibiting the binding of viral protein ORF8 to porphyrins “to a certain extent” [24]. Favipiravir could prohibit ORF7a from binding to (free) porphyrin in addition to preventing the virus from entering the host cells [24].

Starting from this theory, we investigate the host interactome for potential drug repurposing candidates. In CoVex, we selected the viral proteins NSP1-16, ORF3a, ORF7a, ORF8 and ORF10 as seeds and expanded the network to all of their host partner proteins. Notably, we see that NSP7 may bind to Cytochrome b5 reductase (CYB5R), which converts methemoglobin to hemoglobin (oxygen-transporting $Fe^{2+}$ hemoglobin to non-oxygen-transporting $Fe^{3+}$ hemoglobin). Cytochrome b5 reductase is involved in the transfer of reducing equivalents from the physiological electron donor, NADH, via an FAD domain to the small molecules of cytochrome b5. It is also heavily involved in many oxidation and reduction reactions, such as the reduction of methemoglobin to hemoglobin [25–27]. In addition, we see that ORF3a binds to HMOX1 (Heme oxygenase 1), which might lead to interference with heme degradation. We continued with all 238 host interactors as new seeds and executed KeyPathwayMiner (Table S4) to investigate the host interactome for proteins that connect the selected virus-host interactions proteins. We discovered five new drug targets (the proteins APP, XPO1, TRIM25, HSCB, FBXO6) for which we extracted 20 drug candidates using the closeness centrality measure and only screening for currently approved drugs. In the resulting ranked list, we rediscovered Chloroquine as well as Deferoxamine, both of which are currently in clinical trial or discussed in the literature as candidate drugs for COVID-19 treatment. Note that Deferoxamine is widely used for the treatment of Thalassemia and as a chelator of ferric ion in disorders of iron overload [28]. In addition to these two drugs, we find Methylene blue, a drug that is approved by the FDA for the treatment of methemoglobinemia. Note that the evidence for a methemoglobinemia caused by SARS-CoV-2 is anecdotal (no reports on abnormal methemoglobin levels, iron metabolism markers, etc. exist) and we used this hypothesis to illustrate a potential hybrid level starting point for the network medicine investigation of a hypothesis using CoVex.

**Table S4 - Algorithms and parameters used in application scenario d**

| | Algorithm | |
|---|---|---|
| | KeyPathwayMiner | Closeness centrality |
| K | 5 | N.A. |
| Result Size | N.A. | 20 |
| Maximum Degree | N.A. | DISABLED |
| Hub Penalty | N.A. | 0.5 |
| Include indirect drugs | N.A. | FALSE |
| Include non-approved drugs | N.A. | FALSE |

**Table S5 - All drugs in clinical trials targeting host proteins sorted by classification.** We compiled a list of all drugs targeting human proteins currently registered in clinical governmental trial databases (see full paper).

| Drug Classification | Class ID | Drug names |
|---|---|---|
| Immunostimulants | L03 | Sargramostim,Peginterferon alfa-2a,Peginterferon beta-1a |
| Immunosuppressants | L04 | Eculizumab,Thalidomide,Pirfenidone,Anakinra,Fingolimod,Tocilizumab,Sarilumab, Ixekizumab,Siltuximab,Leflunomide,Emapalumab,Tofacitinib,Adalimumab,Baricitinib |
| Antineoplastic agents | L01 | Ruxolitinib,Bevacizumab |
| Gynecological Anti Infectives And Antiseptics | G01 | Darunavir,Ascorbic acid,Sildenafil |
| Vitamins | A11 | Calcitriol,Ascorbic acid |
| Ophthalmologicals | S01 | Heparin,Dexamethasone,Acetylcysteine,Ascorbic acid,Methylprednisolone,Azithromycin |
| Antipsoriatics | D05 | Calcitriol |
| Lipid Modifying Agents | C10 | Pravastatin |
| Antiobesity Preparations;excl.Diet Products | A08 | Phentermine |
| Antivirals For Systemic Use | J05 | Atazanavir,Ribavirin,Oseltamivir,Cobicistat,Sofosbuvir,Ritonavir,Darunavir,Lopinavir |
| Urologicals | G04 | Sildenafil |
| Antibacterials For Systemic Use | J01 | Azithromycin |
| Other Respiratory System Products | R07 | Nitric Oxide |
| Antiprotozoals | P01 | Suramin,Hydroxychloroquine,Artenimol,Chloroquine |
| Psycholeptics | N05 | Dexmedetomidine |
| Agents Acting On The Renin-angiotensin system | C09 | Losartan |
| Allothertherapeuticproducts | V03 | Oxygen,Deferoxamine,Acetylcysteine,Cobicistat |
| Other Gynecologicals | G02 | Naproxen |
| Topical Products For Joint And Muscular Pain | M02 | Naproxen |
| Anti Inflammatory And Antirheumatic Products | M01 | Naproxen |
| Anesthetics | N01 | Sevoflurane,Propofol |
| Anthelmintics | P02 | Levamisole |
| Anti-acne preparations | D10 | Dexamethasone,Methylprednisolone |
| Corticosteroids For Systemic Use | H02 | Dexamethasone,Methylprednisolone |
| Corticosteroids;dermatological preparations | D07 | Dexamethasone,Methylprednisolone |
| Antithromboticagents | B01 | Heparin,Enoxaparin,Dipyridamole |
| Vasoprotectives | C05 | Heparin,Dexamethasone |
| Nasal Preparations | R01 | Ciclesonide,Dexamethasone |
| Ophthalmological And Cytological Preparations | S03 | Dexamethasone |
| Otologicals | S02 | Dexamethasone |
| Stomatological Preparations | A01 | Dexamethasone |
| Antigout Preparations | M04 | Colchicine |
| Drugs for obstructive airway diseases | R03 | Ciclesonide,Mepolizumab |
| Coughandcoldpreparations | R05 | Acetylcysteine,Bromhexine,Noscapine |
| Cardiac therapy | C01 | Angiotensin II |
| Anti Hemorrhagic | B02 | Camostat |
| Bile And Liver Therapy | A05 | Glycyrrhizic acid |